\documentclass[conference]{IEEEtran}
\IEEEoverridecommandlockouts
\usepackage{cite}
\usepackage{dblfloatfix}
\usepackage{placeins}
\usepackage{amsmath,amssymb,amsfonts}
\usepackage{algorithmic}
\usepackage{graphicx}
\usepackage{textcomp}
\usepackage{xcolor}
\usepackage[utf8]{inputenc}
\usepackage{longtable}
\usepackage{booktabs}
\usepackage{array}
\usepackage{url}
\usepackage[table]{xcolor}
\usepackage{geometry}
\usepackage{amsmath,amssymb}
\usepackage{multirow}
\usepackage{graphicx}
\usepackage{arydshln}
\usepackage[square,sort,comma,numbers]{natbib}
\usepackage{colortbl}
\definecolor{lightgray}{gray}{0.92}
\definecolor{midgray}{gray}{0.85}
\definecolor{darkgray}{gray}{0.78}

\definecolor{lightblue}{RGB}{220,230,242}
\definecolor{midblue}{RGB}{200,215,235}
\definecolor{darkblue}{RGB}{180,200,225}

\newif\ifcomments
\commentstrue

\def\BibTeX{{\rm B\kern-.05em{\sc i\kern-.025em b}\kern-.08em
    T\kern-.1667em\lower.7ex\hbox{E}\kern-.125emX}}
\begin{document}
\title{Attribution-Driven Explainable Intrusion Detection with Encoder-Based Large Language Models}

\author{\IEEEauthorblockN{1\textsuperscript{st} Umesh Biswas\textsuperscript{*}, 2\textsuperscript{nd} Shafqat Hasan\textsuperscript{*}, 3\textsuperscript{rd} Syed Mohammed Farhan\textsuperscript{*}, 4\textsuperscript{th} Nisha Pillai, 5\textsuperscript{th} Charan Gudla}
\IEEEauthorblockA{\textit{Computer Science \& Engineering, Mississippi State University}\\
{ \{ucb5\ \textbar \ sh2924\ \textbar \  sm3955\}@msstate.edu, \{pillai\ \textbar \   gudla\}@cse.msstate.edu }}

\thanks{\textsuperscript{*}These authors contributed equally to this work:
Umesh Biswas, Shafqat Hasan, Syed Mohammed Farhan.}
}


\maketitle

\begin{abstract}
Software-Defined Networking (SDN) improves network flexibility but also increases the need for reliable and interpretable intrusion detection. Large Language Models (LLMs) have recently been explored for cybersecurity tasks due to their strong representation learning capabilities; however, their lack of transparency limits their practical adoption in security-critical environments. Understanding how LLMs make decisions is therefore essential. This paper presents an attribution-driven analysis of encoder-based LLMs for network intrusion detection using flow-level traffic features. Attribution analysis demonstrates that model decisions are driven by meaningful traffic behavior patterns, improving transparency and trust in transformer-based SDN intrusion detection. These patterns align with established intrusion detection principles, indicating that LLMs learn attack behavior from traffic dynamics. This work demonstrates the value of attribution methods for validating and trusting LLM-based security analysis. 
\end{abstract}


\section{Introduction}
Software-Defined Networking (SDN) has become a widely adopted architecture in modern networks due to its flexibility, centralized control, and support for dynamic traffic management \cite{priyadarsini2021software}. However, this same flexibility also introduces new security challenges, as SDN controllers must monitor and react to large volumes of network traffic in real time. Intrusion Detection Systems (IDS) play a critical role in identifying malicious behavior in such environments, and recent research has explored the use of machine learning to improve detection accuracy using flow-level network features \cite{alzahrani2021designing}. 

Large Language Models (LLMs) have recently gained attention beyond natural language processing and are increasingly explored for structured data analysis tasks, including cybersecurity \cite{xu2024large,sarker2024generative,kasri2025vulnerability}. Their strong representation learning capability makes them attractive candidates for intrusion detection and security analytics. However, most existing studies focus primarily on detection performance, treating LLMs as black-box classifiers \cite{mehrotra2024tree,zeng2024dald,wu2025survey}. In security-critical systems, this lack of transparency is problematic, as operators must understand why a model flags traffic as malicious in order to trust and act on its decisions. 

Explainability has therefore become an essential requirement for intrusion detection in SDN environments, motivating the use of attribution methods such as Integrated Gradients (IG) to analyze LLM decision behavior. While several explainable AI techniques exist, there is limited work that systematically examines how LLMs reason over flow-level network features and whether their decisions align with well-known intrusion detection principles \cite{mohale2025evaluating,abou2022should}. In particular, it remains unclear whether different LLM architectures rely on similar security-relevant features or whether their predictions are driven by unstable or model-specific patterns. 

\begin{figure*}[t]
\includegraphics[width=0.98\textwidth]{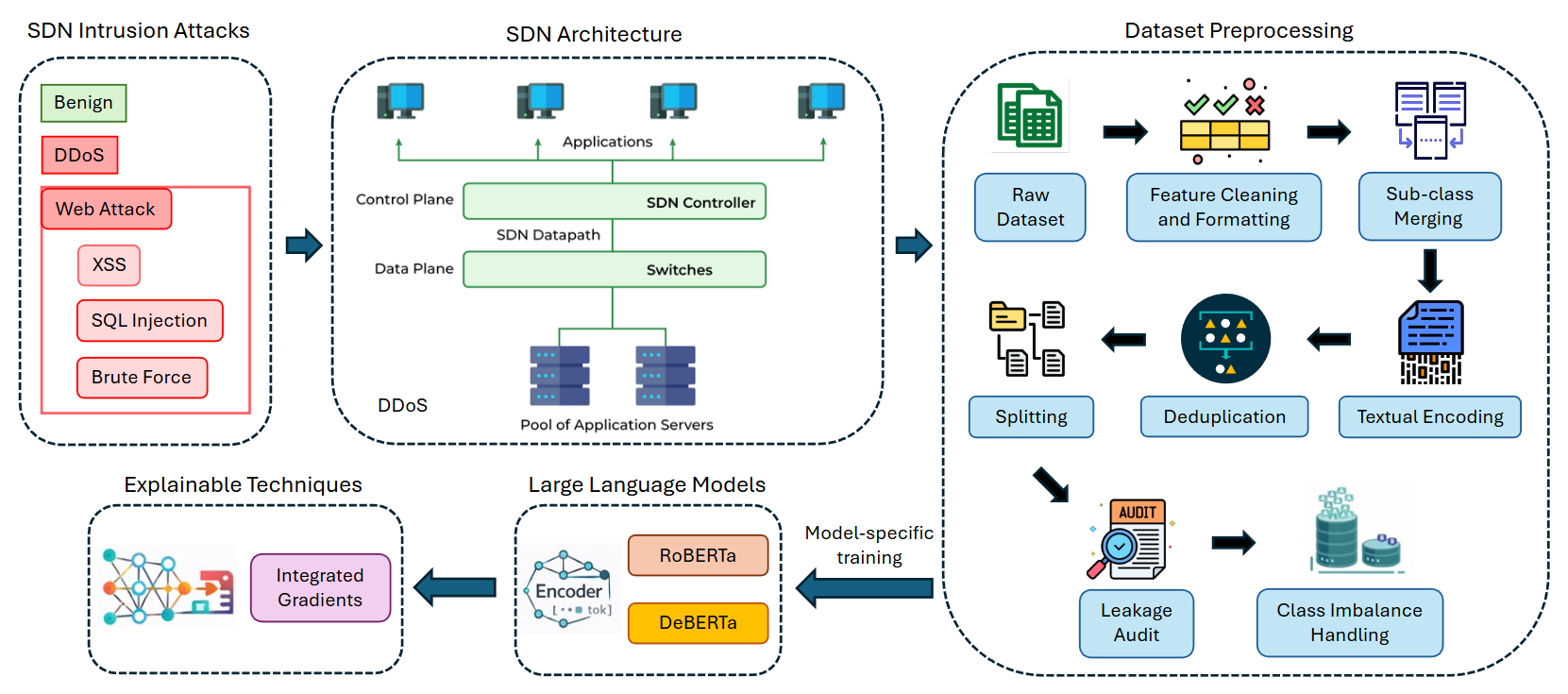}
\vspace{-15pt}
\caption{Overview of the proposed SDN intrusion detection framework, illustrating dataset preprocessing, textual encoding of flow features, COARSE label configuration, encoder-based LLM classification using RoBERTa and DeBERTa, and feature-aligned explainability via Integrated Gradients. }
\label{fig:approach}
\end{figure*}

This work addresses this gap by presenting an attribution-driven analysis of encoder-based LLMs for network intrusion detection (Figure~\ref{fig:approach}). Rather than proposing a new detection system or benchmarking a large number of models, our focus is on understanding and validating model behavior. This study focuses on validating and interpreting LLM decision behavior for SDN intrusion detection rather than proposing a new IDS model, emphasizing explainability, trust, and alignment with established network-security principles. We select RoBERTa and DeBERTa as representative encoder-based LLMs due to their widespread use and architectural differences, and apply IG to identify the flow-level features that most influence their predictions. 

Using the CICIDS2017~\cite{cic_ids_2017} dataset in an SDN context, we analyze how both models attribute importance to key traffic characteristics such as flow duration, packet rate, and inter-arrival timing. By comparing attribution patterns across traffic classes, we examine the extent to which LLM predictions are consistent, interpretable, and aligned with known attack behaviors. Our results show that, despite architectural differences, both models rely on a common set of security-relevant features, while exhibiting only minor variations in secondary cues. 

The contributions of this study are threefold. First, we provide an explainability-focused evaluation framework for analyzing LLM behavior in intrusion detection tasks. Second, we demonstrate how attribution analysis can reveal both shared reasoning patterns and model-specific preferences in encoder-based LLMs. Finally, we show that Integrated Gradients offers a practical tool for validating whether LLM predictions are grounded in meaningful network behavior, supporting their use as explainable and trustworthy components in SDN-based intrusion detection systems.


\section{Related Work}\label{sec-related-work}
\noindent In this section, we review existing work on the application of traditional ML and LLMs to network anomaly detection, with a particular focus on SDN use cases, efficiency optimization, and explainability.

Traditional ML techniques have been widely studied for network intrusion detection. Early approaches primarily rely on supervised learning models such as decision trees \cite{kotsiantis2013decision}, support vector machines (SVM) \cite{steinwart2008support}, k-nearest neighbors (KNN) \cite{guo2003knn}, random forests (RF) \cite{breiman2001random}, and gradient boosting classifiers \cite{natekin2013gradient} applied to detect malicious network traffic from handcrafted numerical features extracted from packet flows or sessions \cite{sommer2010outside,buczak2016survey}. Several studies have explored the role of feature representation and dimensionality reduction in improving machine-learning-based intrusion detection for SDN environments~\cite{berei2024machine}. These approaches emphasize that the effectiveness of classical classifiers is strongly influenced by how flow-level features are selected, transformed, and represented, rather than by model choice alone. Such findings highlight the dependence of traditional ML-based intrusion detection systems on carefully engineered features and well-defined traffic conditions. Early intrusion detection in SDN has also been examined using flow-based features derived from a small number of packets per flow~\cite{towhid2023early}. This work highlights that limited traffic observations can affect the reliability of learned flow statistics in practical deployment scenarios. Despite their strong detection performance in SDN environments, traditional ML models offer limited explainability, typically limited to feature importance scores or heuristic analyses that provide only high-level insight into complex attack behaviors and decision rationale.


Large Language Model (LLM) based approaches have been explored for intrusion detection in SDN, focusing on detection capability, efficiency, and interpretability. GPT-3 models have been adapted for anomaly detection~\cite{balasubramanian2023transformer}, while GPT-4 has been examined using in-context learning with limited labeled traffic data~\cite{zhang2024large}. Transformer-based~\cite{vaswani2017attention} and BERT-based~\cite{devlin2019bert} architectures have been adapted directly to SDN datasets. Hybrid architectures combining Transformers and CNNs achieve strong DDoS detection performance on CICDDoS2019~\cite{wang2021ddostc}, while encoder-only Transformer-based IDS models report high accuracy on the SDN dataset \cite{ataa2024intrusion}. Fine-tuned BERT models applied to textualized SDN flow features further enable the detection of both known and zero-day attacks \cite{swileh2025unseen}. To satisfy SDN’s real-time constraints, optimization techniques such as INT8 and low-bit quantization are employed to significantly reduce memory usage and inference latency with minimal accuracy loss \cite{adjewa2024efficient,rajapaksha2023improving}. Beyond detection, LLMs have also been examined for explainability, where prompted models generate human-readable rationales for malicious SDN flows \cite{houssel2026ex,yang2025large}. Unlike earlier approaches, this work employs encoder-only LLM's to learn representations directly from serialized flow-level SDN traffic features.


Explainable artificial intelligence \cite{yu2023temporal,arous2025llm,bilal2025llms} has gained significant attention in intrusion detection to improve model transparency and trustworthiness. Popular explanation techniques include feature importance measures, LIME~\cite{ribeiro2016should,salih2025perspective}, SHAP~\cite{lundberg2017unified,lodh2025lightweight}, and gradient-based attribution methods such as Integrated Gradients (IG)~\cite{sundararajan2017axiomatic}. These techniques aim to identify which input features contribute most to a model’s prediction, either globally or at the instance level. Among these methods, IG is particularly well-suited for transformer-based language models due to its axiomatic guarantees. This work uses IG to examine class-level attribution patterns across traffic categories, providing insight into model reasoning beyond instance-level explanations. 

\section{Proposed Approach}
Our goal is to study an encoder-based LLM for SDN intrusion detection in a controlled framework that enables strong detection performance while supporting feature-aligned, human-interpretable explanations. To this end, we combine a structured transformation of SDN flow features with encoder-based language models and IG for attribution-driven explainability, as illustrated in Figure~\ref{fig:approach}.

\subsection{Data Representation}\label{sub-data-representation}
We consider a labeled SDN traffic dataset \cite{cic_ids_2017} consisting of flow-level statistical features extracted from packet traces. Each flow is characterized by numerical attributes such as packet counts, duration statistics, byte volumes, and TCP flag information, and is labeled as benign or malicious across multiple attack classes, as shown in Table~\ref{tab:dataset_distribution}.

\begin{table}[htbp]
\centering
\caption{Dataset Class Distribution}
\label{tab:dataset_distribution}
\renewcommand{\arraystretch}{1.15}
\begin{tabular}{lc}
\toprule
\textbf{Class} & \textbf{Number of Samples} \\
\midrule
\rowcolor{midblue}
BENIGN & 243,212 \\
\rowcolor{lightblue}
DDoS & 121,606 \\
\rowcolor{darkblue}
Web Attack – Brute Force & 1,408 \\
\rowcolor{midblue}
Web Attack – XSS & 624 \\
\rowcolor{darkblue}
Web Attack – SQL Injection & 21 \\
\bottomrule
\end{tabular}
\end{table}

While traditional machine learning models operate directly on numerical feature vectors, LLMs require sequential textual input. Rather than relying on handcrafted embeddings or ad-hoc feature encoders, we adopt a deterministic text-based transformation that preserves the semantic identity of each SDN feature.

Let $x = (x_1, x_2, \ldots, x_d)$ denote a flow-level feature vector extracted from the SDN dataset, where $d$ is the total number of numerical traffic features per flow. We define $s(x)$ as the ordered concatenation of feature--value tokens:
\begin{align*}    
s(x) = \bigoplus_{i=1}^{d} \texttt{``Feature}_i\texttt{ is }x_i\texttt{''}.
\end{align*}

This design enforces a consistent feature order across samples, enabling reliable tokenization and, critically, a direct mapping between input tokens and original SDN features during explainability analysis.

\subsection{Model Architecture}
The proposed framework employs a single-head encoder-based architecture for Coarse 3-way intrusion detection. The encoder output is connected to a classification head trained to predict the top-level intrusion labels: Benign, DDoS, and Web Attack, where all web-based attack subclasses (Brute Force, XSS, and SQL Injection) are merged into a single Web Attack category.

This single-head formulation represents the most direct mapping from flow-level traffic features to intrusion classes and serves as the sole detection architecture evaluated in this work. All performance analysis and attribution-based explainability results are reported with respect to this COARSE classification output.

\subsubsection{LLM Architectures}
To study the impact of encoder design on both detection performance and explainability, we employ two representative encoder-based language models: RoBERTa and DeBERTa.

RoBERTa \cite{liu2019roberta} is a 12-layer encoder-only Transformer trained using masked language modeling with optimized pretraining strategies, including larger batch sizes and longer training schedules. Its strong bidirectional contextual modeling makes it well-suited for classification tasks that require holistic reasoning over structured textual inputs, such as textualized SDN flow descriptions.

DeBERTa \cite{he2020deberta} extends the Transformer architecture by disentangling content and positional information within the attention mechanism and introducing an enhanced mask decoder during pretraining. These design choices improve representational expressiveness and generalization, particularly in complex classification settings. By comparing DeBERTa with RoBERTa, we examine whether richer contextual representations translate into improved detection performance and clearer attribution patterns in SDN intrusion detection.

\subsubsection{Fine-Tuning Strategy}
Each textualized flow record is tokenized using the corresponding model tokenizer and padded to a fixed sequence length. To prevent data leakage, duplicate and near-duplicate samples are removed prior to dataset splitting. The dataset is then partitioned into training, validation, and test sets using stratified sampling to preserve class distributions.

SDN intrusion datasets are typically highly imbalanced, with benign traffic dominating the sample distribution. To mitigate this issue, we apply class-weighted cross-entropy loss during training, with class weights set proportional to $1/\sqrt{n_c}$ (with clipping). This places additional emphasis on the minority Web Attack class while maintaining stable optimization. No oversampling is applied to the validation or test sets, ensuring that evaluation metrics reflect realistic deployment conditions.

\begin{table*}[t]
\centering
\caption{Per-class precision (P), recall (R), and F1-score for the merged Web Attack setting (COARSE 3-way).}
\label{tab:merged_perclass_overall_f1}
\renewcommand{\arraystretch}{1.15}
\setlength{\tabcolsep}{6pt}
\begin{tabular}{l|ccc|ccc|ccc|cc}
\toprule
\textbf{Model} &
\multicolumn{3}{c|}{\textbf{BENIGN}} &
\multicolumn{3}{c|}{\textbf{DDoS}} &
\multicolumn{3}{c|}{\textbf{Web Attack}} &
\textbf{F1$_{\text{weighted}}$} &
\textbf{F1$_{\text{macro}}$} \\
& P & R & F1 & P & R & F1 & P & R & F1 & & \\
\midrule
\rowcolor{midblue}
DeBERTa\_Merged &
0.9991 & 0.9998 & 0.9995 &
0.9999 & 0.9989 & 0.9994 &
0.9826 & 0.9611 & 0.9717 &
0.9993 & 0.9902 \\

\rowcolor{lightblue}
RoBERTa\_Merged &
0.9988 & 0.9991 & 0.9989 &
0.9998 & 0.9990 & 0.9994 &
0.9065 & 0.9197 & 0.9130 &
0.9986 & 0.9704 \\
\bottomrule
\end{tabular}
\end{table*}

\subsection{Integrated Gradients for LLM Explainability}\label{sub-IG}
While accurate classification is essential, SDN operators also require interpretable explanations to understand why traffic is flagged as malicious. 
To learn the explanations, we employ Integrated Gradients (IG), a gradient-based attribution method with strong theoretical guarantees, including completeness and implementation invariance. IG quantifies the contribution of each input component to a model's output by integrating gradients along a straight-line path from a baseline input to the actual input.

For transformer-based models, IG is computed with respect to the input token embeddings. Token-level attribution scores are obtained by aggregating IG values (see Equation \ref{equ-1}) across embedding dimensions:
\begin{align}\label{equ-1}
\mathrm{IG}_i(x) = (x_i - x_i^{\prime}) 
\int_{0}^{1} 
\frac{\partial F\!\left(x^{\prime} + \alpha (x - x^{\prime})\right)}
{\partial x_i}
\, d\alpha,
\end{align}
where $F(\cdot)$ denotes the model output for the target class, $x$ is the input, $x^{\prime}$ is a baseline input, and $\mathrm{IG}_i(x)$ measures the attribution of the $i$-th input feature.

Due to the deterministic textual encoding of SDN flow features (Section~\ref{sub-data-representation}), token-level IG attributions can be directly mapped back to original flow-level features. This enables feature-aligned explanations grounded in measurable network characteristics, allowing SDN operators to identify which traffic attributes most influence intrusion decisions and improving transparency and trust.


\section{Experiments \& Evaluation}
Experiments were conducted across two comparable computing environments. Primary large-scale runs were performed on a Dell workstation equipped with an Intel Xeon w9-3495X CPU and an NVIDIA RTX 6000 Ada GPU, while supplementary experiments were executed on a local workstation with an Intel Core i9 CPU and an NVIDIA GeForce RTX 4090 GPU (16 GB VRAM). All systems ran Windows 11 with CUDA support (CUDA 11.8--12.4). The software stack used Python (v3.12--3.13) and PyTorch (v2.2--2.6) \cite{paszke2019pytorch}. Model training and evaluation were implemented using the Hugging Face ecosystem \cite{wolf2020transformers}, including Transformers, Datasets, and PEFT for parameter-efficient fine-tuning. Optimization leveraged the BitsAndBytes library where applicable. Interpretability analyses were performed using Captum, and standard scientific libraries (NumPy, Pandas, Scikit-learn, Matplotlib, Seaborn, and TQDM) were used for data preprocessing, evaluation, and visualization.

\subsection{Dataset Preparation and Evaluation Protocol}
The SDN intrusion dataset is highly imbalanced, with benign and DDoS traffic comprising the majority of samples, while web-based attacks appear far less frequently. To focus on the top-level intrusion detection task studied in this paper, we evaluate a Coarse 3-way label space: Benign, DDoS, and Web Attack. The three web-based attack subclasses (Brute Force, XSS, and SQL Injection) are merged into a single Web Attack class. To prevent train-test leakage, we applied global deduplication using a SHA1 hash over the serialized feature-to-text representation. This reduced the dataset from 1{,}188{,}333 samples to 366{,}870 unique samples. After merging and deduplication, the class distribution was Benign: 243{,}211, DDoS: 121{,}606, and Web Attack: 2{,}053. A leak-safe 70:10:20 train-validation-test split was then applied while preserving class proportions. Overlap audits confirmed zero overlap between training, validation, and test splits. Given the severe class imbalance, overall accuracy alone can be misleading. Therefore, we report macro-averaged F1 as the primary evaluation metric, alongside accuracy and weighted F1. In addition, we report per-class precision, recall, and F1 to explicitly highlight performance on the minority Web Attack class.

\subsection{Models and Training Configurations}
We evaluate two encoder-based LLMs: RoBERTa and DeBERTa, both trained using a single-head architecture for Coarse 3-way classification. Each model is fine-tuned directly on the merged label space (Benign/DDoS/Web Attack), providing a consistent and direct comparison between architectures. To mitigate class imbalance during training, we employ a class-weighted cross-entropy objective. Class weights are set proportional to $1/\sqrt{n_c}$ with clipping, placing additional emphasis on the minority Web Attack class while maintaining stable optimization. All other training settings, input representations, and optimization procedures are kept identical across models to ensure a fair comparison.

\FloatBarrier
\subsection{Results (COARSE 3-way)}

Table~\ref{tab:merged_perclass_overall_f1} summarizes per-class precision, recall, and F1, along with weighted and macro F1 scores, for the merged Web Attack setting. Both models achieve near-perfect overall accuracy due to the dominance of Benign and DDoS traffic. However, macro-level performance reveals meaningful differences in minority-class detection. DeBERTa\_Merged achieves a macro-F1 of 0.9902 and a Web Attack F1 of 0.9717, indicating strong balance between precision and recall on the minority class. In contrast, RoBERTa\_Merged attains a macro-F1 of 0.9704 and a Web Attack F1 of 0.9130, reflecting comparatively lower sensitivity to web-based attacks.

\begin{figure*}[t]
\centering
\includegraphics[width=0.95\textwidth,
  clip,
  trim=0 0 0 0.8cm]{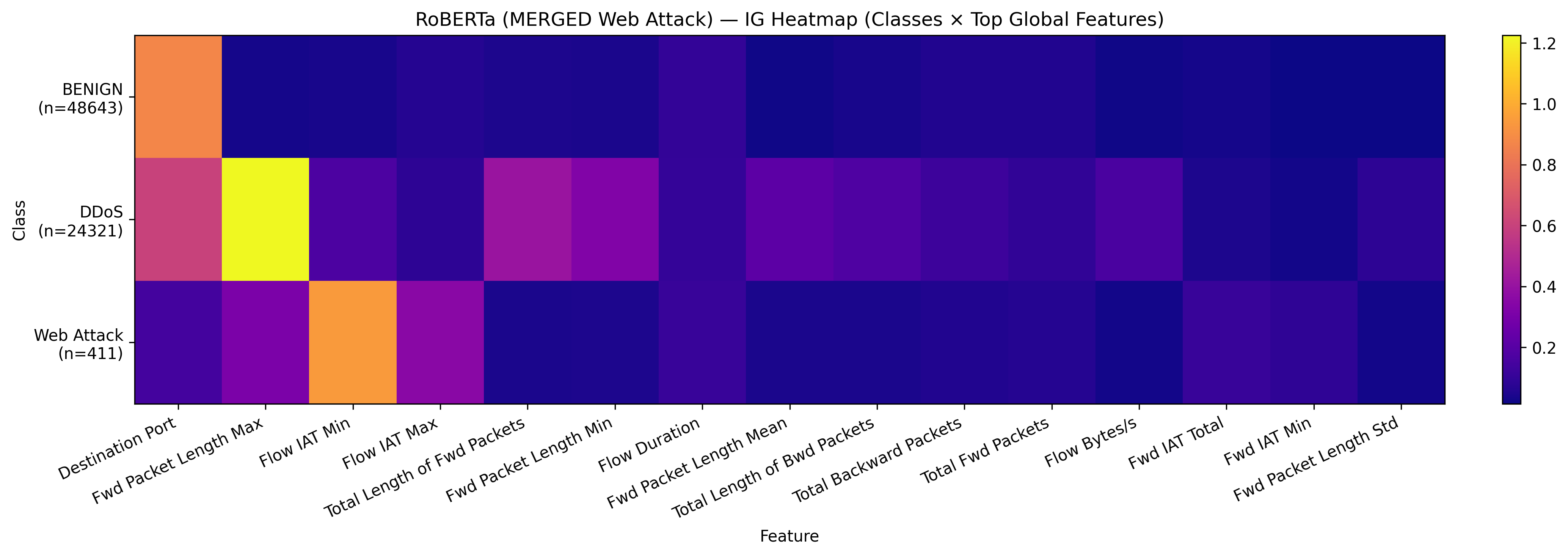}
\caption{Integrated Gradients (IG) heatmap for the RoBERTa-based SDN intrusion detection model. Rows correspond to the COARSE intrusion classes (BENIGN, DDoS, Web Attack), and columns denote the top global SDN flow features ranked by mean absolute attribution.}
\label{fig:ig_roberta_merged}
\end{figure*}

\subsection{Per-Class Performance Analysis}
Per-class results in Table~\ref{tab:merged_perclass_overall_f1} show that both models classify Benign and DDoS traffic almost perfectly, with precision and recall exceeding 0.998 in all cases. The primary source of performance variation lies in the \texttt{Web Attack} class. DeBERTa\_Merged achieves higher precision (0.9826) and recall (0.9611) for Web Attacks, resulting in a strong F1 score of 0.9717. RoBERTa\_Merged, while still effective, exhibits lower precision (0.9065) and recall (0.9197), yielding a Web Attack F1 of 0.9130. These results indicate that DeBERTa provides more reliable detection of minority web-based attacks, which directly contributes to its higher macro-F1 score. Overall, the results demonstrate that while both encoder-based LLMs perform extremely well on majority traffic, DeBERTa offers improved robustness and sensitivity for minority intrusion detection in the merged Coarse 3-way setting.

\section{Discussion}

\begin{figure*}[t]
\centering
\includegraphics[width=0.95\textwidth,
  clip,
  trim=0 0 0 0.8cm]{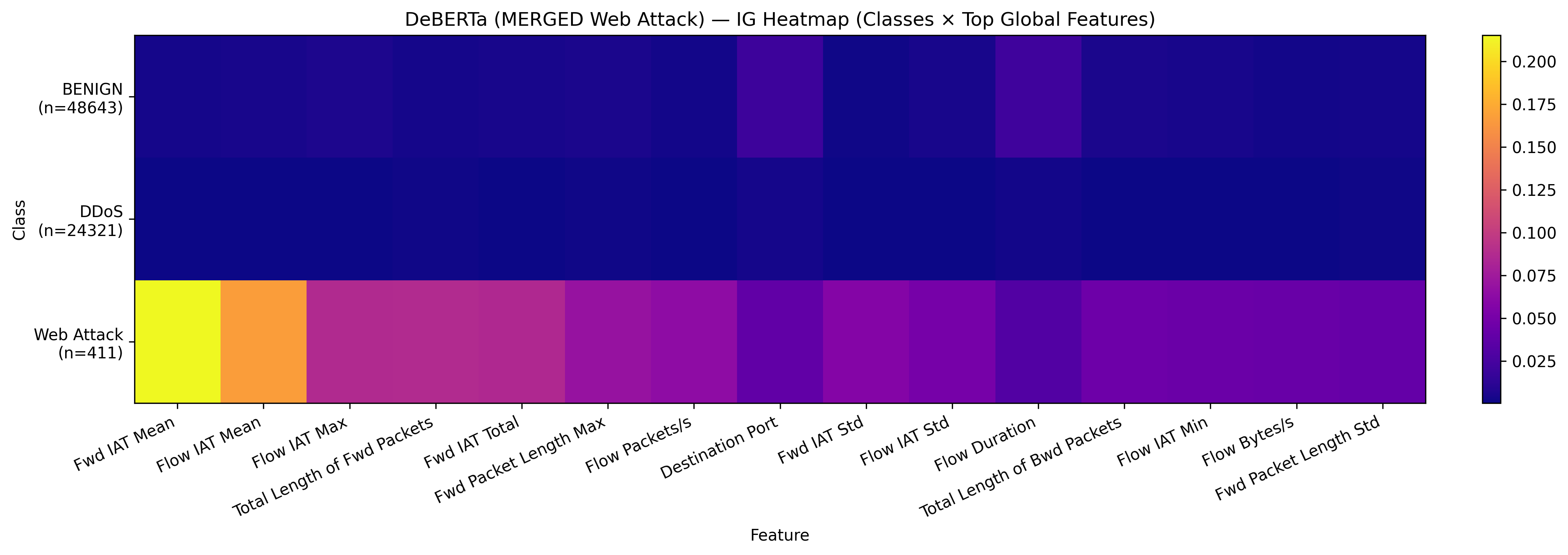}
\caption{Integrated Gradients (IG) heatmap for the DeBERTa-based SDN intrusion detection model. Rows correspond to the COARSE intrusion classes (BENIGN, DDoS, Web Attack), and columns denote the top global SDN flow features ranked by mean absolute attribution.}
\label{fig:ig_deberta_merged}
\end{figure*}

This section interprets the Integrated Gradients (IG) attribution heatmaps for RoBERTa (Figure~\ref{fig:ig_roberta_merged}) and DeBERTa (Figure~\ref{fig:ig_deberta_merged}) under the merged Web Attack setting. The objective is to understand how both encoder-based models distinguish between BENIGN, DDoS, and Web Attack traffic, and how architectural differences influence feature usage. Unlike performance metrics alone, attribution analysis reveals why a model makes a decision, which is particularly important for security-sensitive applications such as software-defined networking (SDN).

Throughout this discussion, higher attribution corresponds to lighter colors (yellow/orange), while lower attribution corresponds to darker colors (blue/purple), as indicated by the color bars in the heatmaps.

\subsection{BENIGN Traffic: Absence of Anomalous Structure}

RoBERTa exhibits a strong attribution peak for Destination Port, indicating reliance on service level context when identifying benign traffic. Features such as Total Duration of a Network Flow (Flow Duration) and Total Bytes in Forward Direction (Total Length of Fwd Packets) show only weak or moderate attribution, suggesting that RoBERTa's benign classification is driven primarily by port information rather than detailed behavioral dynamics. While this strategy yields high BENIGN precision and recall (Table \ref{tab:merged_perclass_overall_f1}), it also highlights a potential limitation: reliance on destination port may reflect dataset regularities rather than intrinsic benign behavior. This is precisely the type of insight that attribution analysis is intended to surface.

DeBERTa, in contrast, shows low and distributed attribution across nearly all features, including Destination Transport Layer Port Number (Destination Port). Small contributions appear for Flow Duration and Destination Port, but these are noticeably weaker than in RoBERTa. This suggests that DeBERTa treats BENIGN traffic as a baseline class, characterized by the absence of strong attack-like signals, rather than by the presence of a specific identifying feature. Such behavior is generally considered more robust in intrusion detection, as it reduces dependence on dataset-specific shortcuts. 

The contrast between models is clear and important:
\begin{itemize}
\item RoBERTa identifies BENIGN traffic primarily using Destination Port.
\item DeBERTa identifies BENIGN traffic through weak, distributed evidence, avoiding a single dominant cue.
\end{itemize}
This difference is consistent with the architectural design of DeBERTa, which enables richer contextual reasoning and may contribute to better generalization beyond the training distribution.

\subsection{DDoS Traffic: High-Intensity and Repetitive Behavior}

For the DDoS class, RoBERTa assigns the highest importance to Maximum Forward Packet Length (Fwd Packet Length Max), with additional influence from Destination Transport Layer Port Number (Destination Port) and Total Bytes in Forward Direction (Total Length of Fwd Packets). This pattern suggests that RoBERTa associates DDoS traffic with extreme packet size behavior and volume-related cues. Such patterns are plausible, as many DDoS traces contain repeated packets with consistent sizes or large payloads, depending on the attack type. The moderate importance of Destination Port may reflect the concentration of attack traffic toward a particular service during flooding. 

In DeBERTa, the DDoS row appears largely uniform and low in attribution across the selected top features. Unlike RoBERTa, no single feature stands out strongly. This does not indicate poor performance; rather, it suggests that DeBERTa may rely on a broader set of smaller signals, many of which may not appear in the global top-15 feature list used for the heatmap. Given that DDoS detection performance is near perfect for both models (Table \ref{tab:merged_perclass_overall_f1}), the relatively flat IG pattern for DeBERTa indicates that classification confidence is achieved without heavy reliance on any single dominant feature.

Both models correctly identify DDoS traffic, but they do so differently:

\begin{itemize}
\item RoBERTa emphasizes packet size and volume-related extremes.
\item DeBERTa relies on distributed evidence, likely combining multiple weaker cues.
\end{itemize}

This again reflects a difference between localized versus distributed reasoning strategies.

\subsection{Web Attack Traffic: Timing-Centric Behavioral Signatures}

The Web Attack class exhibits the richest attribution structure and the clearest differences between DeBERTa and RoBERTa. This is expected, as the merged Web Attack category includes heterogeneous behaviors from brute-force, XSS, and SQL injection attacks.

In RoBERTa, the most influential feature for Web Attack is clearly Minimum Inter-Arrival Time of the Flow (Flow IAT Min), which appears as the brightest cell in the Web Attack row. A secondary contribution comes from Maximum Inter-Arrival Time of the Flow (Flow IAT Max). This pattern indicates that RoBERTa strongly associates web attacks with timing extremes: very short gaps between packets (rapid request bursts) and occasional long pauses (waiting for server responses). These behaviors are common in scripted attacks and automated exploitation tools. Other features, such as Total Duration of a Network Flow (Flow Duration) and Total Forward Inter-Arrival Time (Fwd IAT Total), show only modest attribution, suggesting that RoBERTa's decision boundary is largely driven by extreme timing cues rather than overall flow structure.

DeBERTa presents a markedly different picture. The Web Attack row shows moderate-to-high attribution across a wider set of features, including: Mean Forward Inter-Arrival Time (Fwd IAT Mean), Mean Inter-Arrival Time of the Flow (Flow IAT Mean), Maximum Inter-Arrival Time of the Flow (Flow IAT Max), Total Forward Inter-Arrival Time (Fwd IAT Total), Total Bytes in Forward Direction (Total Length of Fwd Packets), Flow Packet Rate (Flow Packets/s), Standard Deviation of Forward Inter-Arrival Time (Fwd IAT Std), Standard Deviation of Flow Inter-Arrival Time (Flow IAT Std), and Total Duration of a Network Flow (Flow Duration). This distribution shows that DeBERTa does not rely on a single extreme feature. Instead, it integrates average timing, timing variability, request intensity, and flow persistence. This richer representation aligns well with the complex and varied nature of web attacks when grouped into a single class.

In summary, RoBERTa focuses on timing extremes, especially Flow IAT Min and DeBERTa integrates multiple complementary timing and flow features. This difference directly corresponds to performance: DeBERTa achieves a substantially higher Web Attack F1 score in the merged setting.

\subsection{Architectural Interpretation}

Destination Transport Layer Port Number (Destination Port) plays a class-dependent role in both models. In RoBERTa, it is highly influential for BENIGN, moderately influential for DDoS, and less influential for Web Attack, whereas in DeBERTa, it appears only as a secondary supporting feature across all classes. This suggests that RoBERTa is more prone to service-level shortcuts, while DeBERTa relies primarily on behavioral patterns. From an SDN security perspective, the latter is generally preferable, as attackers often target the same services used by benign users. 

These attribution patterns are consistent with known architectural differences. DeBERTa's disentangled attention mechanism separates content and positional information, enabling the model to represent relationships among timing statistics more effectively. RoBERTa, using a standard Transformer encoder, appears to prioritize salient, high-contrast signals such as extreme timing values or strongly discriminative ports. This can be effective but may also lead to over-reliance on a small number of features.

\subsection{Implications for Explainable SDN Intrusion Detection}
The combined findings demonstrate that encoder-based LLMs can learn meaningful, interpretable traffic representations from flow-level features alone. Integrated Gradients exposes whether a model relies on robust behavioral signals or potentially brittle shortcuts.

In this study BENIGN traffic is characterized by weak or service-context cues, DDoS traffic is identified through intensity and repetition, and Web attacks are identified primarily through timing structure. These patterns align well with established intrusion detection theory and provide confidence that the learned representations are behaviorally grounded.

\section{Conclusion}

This work investigates the explainability of encoder-only transformer models for SDN intrusion detection using Integrated Gradients. Flow-level traffic features are serialized into textual representations, enabling effective learning with transformer-based architectures. Experimental results show that RoBERTa and DeBERTa achieve strong classification performance while primarily relying on security-relevant behavioral features such as flow duration, packet rates, and inter-arrival time statistics, with service-level context (e.g., destination port) acting as a secondary cue. Attribution analysis reveals both shared and distinct reasoning strategies across models: DeBERTa exhibits a more distributed, behavior-centric reliance on timing and flow dynamics, whereas RoBERTa places greater emphasis on a small set of highly discriminative cues, including destination port and timing extremes. Importantly, these attribution patterns align with established intrusion detection principles, providing transparency into model behavior. Overall, the findings show that transformer-based IDS models can be both accurate and interpretable, supporting their practical deployment in SDN environments where explainability and trust are essential.


\FloatBarrier

\bibliography{reference}

\end{document}